\documentclass[aps,prl,twocolumn,superscriptaddress]{revtex4}

\usepackage{graphicx}
\usepackage{bm}
\usepackage{amsmath}
\usepackage{color}

\begin{document}

\preprint{tsk-etr-v10}

\title{Nonlinear graphene metamaterial}

\author{Andrey E. Nikolaenko}
\affiliation{Optoelectronics Research Centre, University of Southampton, Southampton SO17 1BJ, United Kingdom}

\author{Nikitas Papasimakis}
\affiliation{Optoelectronics Research Centre, University of Southampton, Southampton SO17 1BJ, United Kingdom}
\author{Evangelos Atmatzakis}
\affiliation{Optoelectronics Research Centre, University of Southampton, Southampton SO17 1BJ, United Kingdom}
\author{Zhiqiang Luo}
\affiliation{School of Physical and Mathematical Sciences, Nanyang Technological University 637371 Singapore}

\author{Ze Xiang Shen}
\affiliation{School of Physical and Mathematical Sciences, Nanyang Technological University 637371 Singapore}

\author{Francesco De Angelis}
\affiliation{Italian Institute of Technology, 16163 Genova and the
University of Magna Graecia, 88100 Catanzaro, Italy}

\author{Stuart A. Boden}
\affiliation{School of Electronics and Computer Science, University of Southampton, Southampton SO17 1BJ, United Kingdom}


\author{Enzo Di Fabrizio}
\affiliation{Italian Institute of Technology, 16163 Genova and the
University of Magna Graecia, 88100 Catanzaro, Italy}

\author{Nikolay I. Zheludev}
\affiliation{Optoelectronics Research Centre, University of Southampton, Southampton SO17 1BJ, United Kingdom}

\date{\today}

\begin{abstract}
We demonstrate that the broadband nonlinear optical response of graphene can be resonantly enhanced by more than an order of magnitude through hybridization with a plasmonic metamaterial, while retaining an ultrafast nonlinear response time of $\sim 1$ ps. Transmission modulation close to $\sim 1\%$ is seen at a pump
fluence of $\sim 30~\mu$J/cm$^2$ at the wavelength of $\sim 1.6$ $\mu$m.  This approach allows to engineer and enhance graphene's nonlinearity within a broad wavelength range enabling applications in optical switching, mode-locking and pulse shaping.
\end{abstract}

\maketitle

Graphene, a two dimensional atomic layer of carbon atoms, exhibits remarkable properties that have attracted enormous research interest in recent years \cite{g_geim04,g_geim05,g_geim07}. Due to its unique semi-metallic electronic band structure (see Fig. 1a), charge carriers in graphene behave like massless Dirac fermions providing a model system for investigations of quantum effects \cite{g_geim05, g_kim05}, while its record-high mobility and thermal conductivity suggest that graphene will play a key role in next generation electronics \cite{g_avouris07}. Its optical properties are equally fascinating with a flat absorption profile throughout the near- and mid-infrared parts of the spectrum. This broadband absorption is remarkably strong and even in the case of a single monoatomic layer it amounts to $\sim 2.3\%$. Absorption in graphene is a result of interband transitions, which also lead to nonlinearity: At high excitation energies interband transitions become blocked and absorption decreases. Due to the absence of a band gap, relaxation in graphene occurs through carrier-carrier and carrier-photon interactions allowing ultrafast recovery times at the ps scale \cite{g_spencer08, g_sun08, g_driel09, g_kaindl09, g_huan10,g_luo10,g_breusing11,g_svirko11} . The broadband and ultrafast characteristics of its saturable absorption are already leading to applications of graphene in laser mode-locking \cite{g_ferrari10b, g_tang09}.  However, the application potential of graphene is limited by its weak nonlinearity requiring very high levels of excitation of the order of a few GW/cm$^2$ to engage it. Here we show that combining graphene with a plasmonic metamaterial can lead to more than order of magnitude resonant enhancement of its nonlinear properties while retaining its ultrafast character. Our results allow not only to enhance but also to spectrally tailor the saturable absorption of graphene opening the way for applications in ultrafast all-optical switching.

Hybridization of low dimensional carbon (carbon nanotubes) with plasmonic metamaterials is known to lead to strong interactions and boost the nonlinear response due to the field enhancement provided by the metamaterial resonances, with minor effects, if any, at the recovery times \cite{cntmm,g_me10, cntmm2}. Here, we hybridize graphene with a planar metamaterial consisting of complementary asymmetrically-split rings (ASRs) (see Fig. 1d). Metamaterials of this type are known to support trapped-mode excitations, where weak coupling of plasmonic modes to free-space minimizes radiation losses \cite{asrs}. This manifests as a narrow Fano line in the transmission and reflection spectra and, in the presence of plasmonic losses, it is accompanied by an absorption peak (see Fig. 3) \cite{cntmm}.
\begin{figure} [h!]
\includegraphics[width=0.45\textwidth]{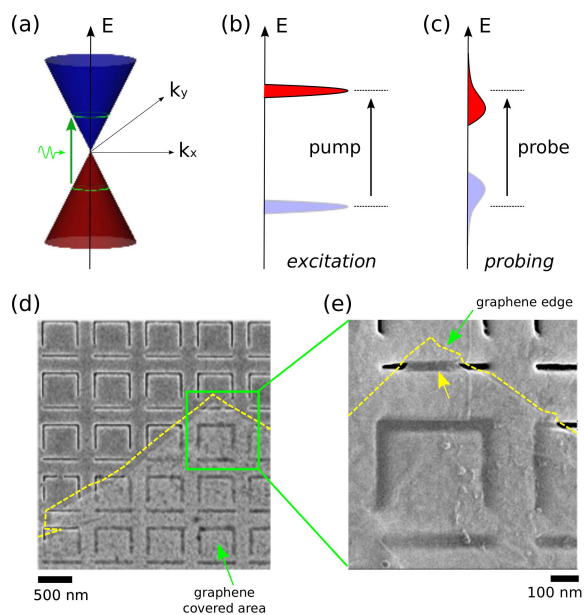}
\caption{(a) The electronic band structure of graphene can be approximated by two cones intersecting at the Dirac point. The pump beam creates a short-lived carrier distribution (b) that blocks the interband transitions induced a few ps later by the probe beam (c), hence leading to an absorption decrease. (d-e) Helium ion microscope images of a metamaterial array partially covered with graphene.}\label{conf}
\end{figure}

The metamaterial structures were fabricated by focused ion beam milling through a 65 nm thick gold film evaporated on a $102$ nm thick Si$_3$N$_4$ membrane. The overall size of the arrays was $22$x$22$ $\mu$m$^2$, while the unit cell was $711$ nm. For wavelengths longer than the unit cell side, the metamaterial arrays do not diffract. The graphene films were grown on polycrystalline Cu foils at temperatures up to $1000^o$C by low pressure CVD process using a mixture of ethylene and hydrogen. After the growth, the graphene layers were separated from the copper foils and transfered on the metamaterial samples. Contrast and Raman spectroscopy and Raman mapping were carried out to assess the quality and uniformity of the CVD graphene \cite{g_me10}.

The nonlinear optical response of graphene and the hybrid graphene metamaterial structure was investigated by single-colour non-collinear pump-probe spectroscopy. The pump and probe beams were provided from the output of a tunable $200$ fs OPO system pumped by a Ti:Sapphire laser operating at $80$ MHz and were focused to diameters of $\sim30~ \mu$m slightly exceeding the size of the metamaterial array. The pump beam was modulated by an acousto-optical modulator (AOM) at $30$ kHz resulting in a pulse duration of $\sim500$ fs. The AOM also shifted the frequency of the pump beam by $29-36$ MHz, which allowed to eliminate completely coherent artefacts in the pump-probe traces, as well as the effects of multi-photon absorption in the gold metamaterial. The pump fluence in spectrally resolved measurements was $29$ $\mu$J/cm$^2$ corresponding to a power density of $58$ MW/cm$^2$. The intensity of the probe pulse was $20$ times weaker than the intensity of the pump pulse.
\begin{figure}
\includegraphics[width=0.5\textwidth]{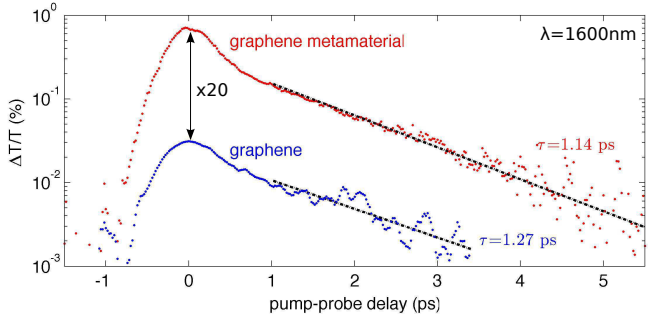}
\caption{Transient relative transmission change as a function of pump-probe time delay for the graphene metamaterial (red dotted line) and graphene on Si$_3$N$_4$ (blue doted line) at a pump fluence of $29~\mu$J/cm$^2$ and wavelength of $1600$ nm. In the case of graphene on Si$_3$N$_4$, the signal reaches noise levels after $3$ ps and hence, subsequent recorded values have been discarded. The black dashed lines represent fits of exponentially decaying functions with characteristic time constants $1.14\pm0.06$ ps and $1.27\pm0.12$ ps, for the graphene metamaterial and graphene on Si$_3$N$_4$, respectively.}\label{ABS}
\end{figure}

The changes in the nonlinear optical properties of graphene due to the hybridization with the plasmonic metamaterial are quantified by the relative transmission change, $\Delta T/T$, where $\Delta T=T-T_0$, while $T$ and $T_0$ represent the graphene metamaterial transmission in the presence and absence of the pump, respectively. The relative transmission changes at the wavelength of the metamaterial resonance ($1600$ nm) are shown in Fig. 2. The graphene-metamaterial structure exhibits a very strong nonlinear response, which exceeds values of $0.7\%$. In contrast, graphene on the Si$_3$N$_4$ substrate, exhibits a much weaker effect of about $0.04\%$. This corresponds to a $20$-fold enhancement of the nonlinear response due to the hybridization with the plasmonic metamaterial.

Importantly, the nonlinear response of graphene, even after hybridization with the plasmonic metamaterial, retains its ultrafast character. The transient nonlinear changes of transmission of graphene follow a bi-exponential dependence with the faster component $\sim 100$ fs (not resolved in our measurements) attributed mainly to carrier-carrier intraband scattering and the slower component $\sim 1$ ps attributed to the carrier-phonon intraband scattering and electron-hole interband recombination processes. The time constant, $\tau$, related to the the slower decay component is estimated by fitting with an exponential the trace of relative transmission change for times longer than $1$ ps. The dynamics of nonlinear changes of transmission of the hybrid graphene-metamaterial structure nearly replicate the ultrafast dynamics of graphene, yielding relaxation times of $\tau=1.14\pm 0.06$ ps and $\tau=1.27\pm 0.12$ ps, respectively, which are indistinguishable within the precision of our measurements. This indicates that the transient nonlinear optical response of graphene metamaterial is governed by relaxation processes in graphene and is not affected by the coupling to the plasmonic resonances.

\begin{figure}
\includegraphics[width=0.5\textwidth]{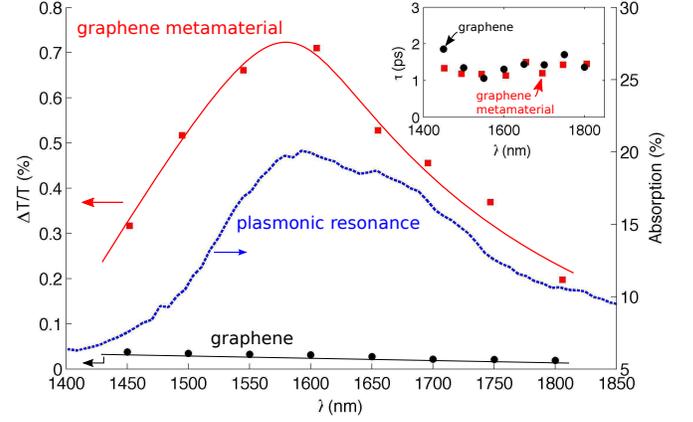}
\caption{(a) Spectral dependence of the nonlinear transmission changes for the graphene metamaterial (red squares) and graphene on Si$_3$N$_4$ (black circles). The red and black lines serve as eye-guides. For comparison, the linear absorption of the graphene metamaterial is also presented (dashed blue line). In the inset, the wavelength dependence of the relaxation time is shown for graphene on Si$_3$N$_4$ (black circles) and the graphene metamaterial (red squares). }\label{SHI}
\end{figure}

The hybridization of graphene with the plasmonic metamaterial provides spectral control over its broadband nonlinear response. This is illustrated in Fig. 3, where the relative transmission change is presented as a function of wavelength for graphene on Si$_3$N$_4$ (black circles) and the graphene-metamaterial (red squares). In the first case, the transmission change takes very low values, smaller than $0.05\%$, and is almost flat, with the exception of a weak decreasing trend with increasing wavelength that is attributed to changes in the interference within the thin Si$_3$N$_4$ membrane due to changes in the absorption of the graphene layer.  On the contrary, the response of the graphene metamaterial is strongly wavelength dependent reaching a peak value of $0.7\%$ at $1600$ nm and rapidly decaying for shorter and longer wavelengths.

The spectral dependence of the nonlinear response follows a trend very similar to the plasmonic metamaterial absorption, represented by the dashed blue line in Fig. 3, which indicates that the mechanism of enhancement of the nonlinear response can be traced in the strong near-fields associated with the plasmonic resonance. Indeed, the absorption peak is a signature of resonantly enhanced plasmonic near fields that can be much higher than the incident field and lead consequently to nonlinear response that is usually observed at higher incident power. At the same time, there is no evidence of further coupling between the metamaterial and the graphene layer, since the time dynamics for both the graphene metamaterial and graphene on Si$_3$N$_4$ over the measurement wavelength range are indistinguishable (inset to Fig. 3) and the ultrafast relaxation times are preserved.

In summary, we show that metamaterials can enhance by more than an order of magnitude the nonlinear response of bare graphene at a specific wavelength within a broad spectral range, without affecting its ultrafast relaxation times. Hybridization with metamaterials allows to spectrally tailor the nonlinear properties of graphene through the resonant features of the metamaterial and we expect that such schemes will facilitate applications in mode-locking, optical switching and pulse shaping.

The authors would like to acknowledge the financial support of the
Engineering and Physical Sciences Research Council (U.K.) and the Royal Society.\\

\end{document}